\newcommand{\mbold}[1]{\mbox{\boldmath$#1$}}

\newcommand{\jpg}{{\em J. Phys. G: Nucl. Part. Phys.} }   

\newcommand{\APNY}{{\em Ann. Phys., NY\/} }

\newcommand{\NP}{{\em Nucl. Phys.} }
\newcommand{\PL}{{\em Phys. Lett.} }
\newcommand{\PR}{{\em Phys. Rev.} }
\newcommand{\PRL}{{\em Phys. Rev. Lett.} }

\newcommand{\ZP}{{\em Z. Phys.} }
\def\sigcom{\sigma_{\pi{\scriptscriptstyle {\rm N}}}}

\def\ubar{\overline u}
\def\dbar{\overline d}
\def\sbar{\overline s}

\documentstyle[12pt,aps]{revtex}
\begin{document}
\title{Does the effective Lagrangian for low-energy QCD scale?}
\author{Michael C. Birse}
\address{Theoretical Physics Group, Department of Physics\\
University of Manchester, Manchester, M13 9PL, U.K.\\}
\maketitle

\begin{abstract}
QCD is not an approximately scale invariant theory. Hence a dilaton field is
not expected to provide a good description of the low-energy dynamics
associated with the gluon condensate. Even if such a field is introduced, it
remains almost unchanged in hadronic matter at normal densities. This is
because the large glueball mass together with the size of the phenomenological
gluon condensate ensure that changes to that condensate are very small at such
densities. Any changes in hadronic masses and decay constants in matter
generated by that condensate will be much smaller that those produced
directly by changes in the quark condensate. Hence masses and decay constants
are not expected to display a universal scaling.
\end{abstract}

It has recently become popular to extend models intended as approximations to
an effective Lagrangian for low-energy QCD by including a dilaton field
[1-8]. This is done in order to make contact with the scale anomaly of QCD, as
discussed by Schechter and others \cite{schdil,gjjs}. Such models have also
been used to justify a universal scaling of all hadron masses and decay
constants in dense matter \cite{brscale}.

The basic ingredient in these models is an extra scalar, isoscalar field, the
dilaton \cite{ellis}, whose vacuum expectation value provides the only scale in
the model. All dimensioned coupling constants are replaced by appropriate
powers of this field multiplied by dimensionless constants. For example, in a
sigma model the pion decay constant becomes a multiple of the dilaton field.
The self-interaction potential for this field, denoted by $\chi$, is taken to
be of the form
$$V(\chi)=a\chi^4+b\chi^4\ln(\chi/\chi_0).  \eqno(1)$$
The first term provides a scale-invariant classical potential which on its own
would lead to a vanishing vacuum expectation value for $\chi$. It would also
leave the dilaton excitations massless, rather like Goldstone bosons. The
second term models the quantum effects responsible for the scale anomaly. It
explicitly breaks scale invariance, driving the vacuum to a nonzero value of
$\chi$ and providing a mass for the dilaton excitations. The single dimensioned
parameter of the model is $\chi_0$, which sets the scale of all other
dimensioned masses and couplings. From a scaling of all dimensioned quantities,
one finds that the $\chi$ field can be related to the trace of the
stress-energy tensor by
$$-4b\chi^4=T_\mu^\mu. \eqno(2)$$
This trace contains all effects which break scale invariance. In QCD it is
dominated by a gluonic contribution which arises from the scale anomaly
\cite{scanom}. The vacuum expectation value of this is given by the gluon
condensate, a phenomenological value for which can be extracted from QCD sum
rules \cite{svz,rry,qssr}.

Such a field can provide a useful description of the low-energy dynamics so
long as the breaking of scale invariance is small. This is analogous to the use
of PCAC and effective chiral Lagrangians to describe the interactions of pions.
There one makes use of the chiral SU(2)$\times$SU(2) symmetry of QCD which is
weakly broken by the current masses of the up and down quarks. The small
explicit symmetry breaking means that pions, although massive, retain much of
their Goldstone boson character. In particular one can use
$$f_\pi m_\pi^2 \mbold\pi=\partial_\mu \mbold A^\mu \eqno(3)$$
to define an interpolating pion field $\pi$. The matrix elements of this field
are dominated by the pion pole, since all other states of the same quantum
numbers lie much higher in energy. This is the basis of partial conservation of
the axial current (PCAC) which can be used to relate interactions of pions with
other particles to the symmetry properties of those particles. These soft pion
theorems are incorporated into the effective chiral Lagrangians which form the
basis of chiral perturbation theory.

Similarly, if the dilaton mass $m_\chi$ were light enough, the relation (2)
could be used to define an interpolating dilaton field by analogy with the pion
field of PCAC (3). This could then be used to obtain ``soft dilaton theorems"
describing the consequences of approximate scale invariance. Lagrangians
including this field and the potential (1) would embody this approximate
symmetry.

QCD is a theory whose Lagrangian is scale invariant at the classical level
(except for the current masses of the quarks) but this invariance is
broken by quantum effects. This breaking is large, as can be seen from the fact
that the lightest scalar glueball, which one might hope to identify with a
dilaton, is estimated to lie at around 1.5 GeV \cite{glueball,morgan}. There
are many other scalar, isoscalar states in the energy range 1--2 GeV and so a
single pole is most unlikely to dominate matrix elements of the stress energy
tensor. Hence an interpolating dilaton field introduced in the above manner is
not a useful ingredient in low-energy effective Lagrangians for QCD.

Moreover, the large mass of the scalar glueball indicates indicates a strong
restoring force against deformations of the gluon condensate. This suggests
that changes to the gluon condensate are likely to be small, both in normal
nuclear matter and in the exotic pionic matter discussed by Mishustin and
Greiner \cite{mishustin}. Hence even if a dilaton field is introduced in
low-energy effective Lagrangians it plays no significant role in the dynamics.
This has been known since such models were first used in the context of hadron
structure [1-8]: significant changes to the gluon condensate are not produced
inside hadrons or normal nuclear matter if realistic values of the glueball
mass and gluon condensate are used. Such small effects as do occur in models
with a dilaton field should not be regarded as reliable estimates because
the scale invariance is so strongly broken.

A clear demonstration of the stiffness of the gluon condensate is provided by
the work of Cohen, Furnstahl and Griegel \cite{cfg}, which uses the trace
anomaly and the Feynman-Hellmann theorem to relate the change in the gluon
condensate to the energy density of hadronic matter. The trace of the stress
energy tensor for QCD is given by the gluonic piece from the scale anomaly plus
terms arising from the current quark masses:
$$T_\mu^\mu=-{9\alpha_s\over 8\pi}G_{\mu\nu}^a G^{a\mu\nu}+m_u\ubar u
+m_d\dbar d+m_s \sbar s,   \eqno(4)$$
where heavy quark contributions have been neglected \cite{cfg}. In the vacuum
this is dominated by the contribution of the gluon condensate $\langle
(\alpha_s/ \pi)G_{\mu\nu}^a G^{a\mu\nu}\rangle\simeq (360\pm 20$ MeV)$^4$
\cite{svz,rry,qssr}.

In stable nuclear matter the pressure vanishes and the change in $T_\mu^\mu$ is
simply the energy density of the matter ${\cal E}$:
$$\langle T_\mu^\mu\rangle_\rho=\langle T_\mu^\mu\rangle_{vac}+{\cal
E}. \eqno(5)$$
Assuming that the change in the nonstrange quark condensate is given by the
leading, model-independent result \cite{druk,cfg} and neglecting the strange
quark content of the proton, the change in the gluon condensate is \cite{cfg}
$$\langle(\alpha_s/ \pi)G_{\mu\nu}^a G^{a\mu\nu}\rangle_\rho-\langle (\alpha_s/
\pi)G_{\mu\nu}^a G^{a\mu\nu}\rangle_{vac}\simeq-{8\over 9}(E-\sigcom)\rho,
\eqno(6)$$
where $E$ denotes the energy per nucleon and $\sigcom$ the pion-nucleon sigma
commutator \cite{sigcom}. The smallness of nuclear binding energies means that
(6) is dominated by the rest masses of the nucleons and so the change in the
gluon condensate is essentially proportional to the baryon density $\rho$.

For normal nuclear matter of density $\rho\simeq 0.17$ fm$^{-3}$, this gives a
change in the gluon condensate of about 150 MeV fm$^{-3}$. This should be
compared with the vacuum gluon condensate of 2200 MeV fm$^{-3}$. Even allowing
for a factor of two uncertainty in this condensate, its change in nuclear
matter is at most a 15\% effect. The fourth root of the condensate, which
corresponds to the change in the dilaton field or the change of scale, is
altered by no more than 4\%. For the pionic droplets studied in
\cite{mishustin} the energy density is even smaller, about 20 MeV fm$^{-3}$ and
so the dilaton field is barely changed.

There are only two ways to get large changes in the gluon condensate at normal
densities relative to its vacuum value. One is to take a value for the vacuum
condensate very much smaller than that the deduced from QCD sum rules. That
would mean rejecting the rather well tested applications of those sum rules to
charmonium \cite{svz,rry,qssr}. The other is to use a $\chi$ field with a very
light mass so that the vacuum is soft in this channel and the response is
large. That would mean returning to the light dilaton idea, even though no such
particle is observed and both lattice calculations and hadron spectroscopy
suggest a scalar glueball mass of about 1.5 GeV \cite{glueball,morgan}. It
would also be inconsistent with observed nuclear binding energies, since
Eq.~(6) provides a connection between these and the change in the gluon
condensate. Neither of these choices seems acceptable.

In summary: QCD is not an approximately scale invariant theory and hence a
dilaton field does not provide a good description of the low-energy dynamics
associated with the gluon condensate (unlike the pion in the context of chiral
dynamics). Moreover, the large glueball mass together with the size of the
phenomenological gluon condensate mean that changes to that condensate are
very small in hadronic matter at normal densities.

A corollary to this is that hadron masses and decay constants do not scale in
matter as suggested by Brown and Rho \cite{brscale}. Any changes in these
quantities are likely to be driven directly by the reduction of the quark
condensate. The model-independent result for the linear dependence of the quark
condensate on density \cite{druk,cfg} shows that large changes in that
condensate can occur independently of any change in the gluon condensate. The
fact that different condensates behave very differently at finite density
should not be too surprising: there are many possible energy scales in matter
which can be constructed from those condensates and the density.

Scaling could only be recovered if one were to use a model where the lightest
scalar meson had a much larger mass than the dilaton, as noted by Kusaka and
Weise \cite{kusaka,kusaka2}.\footnote{The scaling hypothesis leads to hadron
masses which vary as the cube root of the quark condensate. Such a relationship
has also been found in a version of the NJL model without taking a very large
mass for the scalar meson \cite{ripjam}. However in that model the relationship
between the masses and the quark condensate is not a consequence of scaling but
instead arises from the artificial choice of a model involving four-body rather
than two-body forces between the quarks.} The quark condensate would then be
very stiff and would not respond directly to the scalar density of quarks in
matter. Any changes to it could only arise from changes to the gluon
condensate, and hence would be very small for the reasons described above. The
size of the $\pi$N sigma commutator and its associated form factor
\cite{sigcom} indicate that the quark condensate is in fact significantly
deformed in the presence of valence quarks. This can occur even if the
``elementary" scalar meson is heavy because of its strong mixing with the two
pion channel \cite{birse}.

\section*{Acknowledgments}
I am grateful to the ECT$^*$, Trento for its hospitality during the workshop on
Mesons and Baryons in Hadronic Matter. I wish to thank I. Mishustin, J. Wambach
and T. Waindzoch for useful discussions, and J. McGovern for a critical
reading of the manuscript. This work is supported by an SERC Advanced
Fellowship.



\begin{references}
\bibitem{bags}Gomm H, Jain P, Johnson R and Schechter J 1986 \PR {\bf D33}
3476\\
Meissner U-G and Kaiser N 1987 \PR {\bf D35} 2859\\
Meissner U-G, Johnson R, Park N and Schechter J 1988 \PR {\bf D37}
1285
\bibitem{jjs}Jain P, Johnson R and Schechter J 1988 \PR {\bf D38} 1571
\bibitem{eko}Ellis J, Kapusta J I and Olive K A 1991 \NP {\bf B348} 345;
\PL {\bf B273} 123
\bibitem{ripjam}Ripka G and Jaminon M 1992 \APNY {\bf 218} 51\\
Jaminon M and Ripha G 1993 \NP {\bf A564} 551
\bibitem{kusaka}Kusaka K and Weise W 1992 \ZP {\bf A343} 229;
\PL {\bf B288} 6
\bibitem{mbr}Mishustin I, Bondorf J and Rho M 1993 \NP {\bf A555} 215
\bibitem{mishustin}Mishustin I N and Greiner W 1993 \jpg {\bf 19} L101
\bibitem {waindz}Waindzoch T and Wambach J 1994 Private communication
\bibitem{schdil} Schechter J 1980 \PR {\bf D21} 3393\\
Migdal A A and Shifman M A 1982 \PL {\bf B114} 445
\bibitem{gjjs}Gomm H, Jain P, Johnson R and Schechter J 1986 \PR {\bf D33}
801, and references therein
\bibitem{brscale}Brown G E and Rho M 1991 \PRL {\bf 66} 2720
\bibitem{ellis}Ellis J 1970 \NP {\bf B22} 478
\bibitem{scanom}Collins J C, Duncan A and Joglekar S D 1977 \PR {\bf D16}
438\\
Nielsen N K 1977 \NP {\bf B210} 212
\bibitem{svz}Shifman M A, Vainshtein A I and Zakharov V I 1979 \NP {\bf B147}
385, 448
\bibitem{rry}Reinders L J, Rubinstein H and Yazaki S 1985 {\it Phys. Reports}
{\bf 127} 1
\bibitem{qssr}Narison S 1989 {\it QCD spectral sum rules} (Singapore: World
Scientific)
\bibitem{glueball}Bali G S, Schilling K, Hulsebos A, Irving A C, Michael C
and Stephenson P W 1993 \PL {\bf B309} 378
\bibitem{morgan}Morgan D 1993 {\it RAL report} RAL-93-078
\bibitem{cfg}Cohen T D, Furnstahl R J and Griegel D K 1992 \PR {\bf C45} 1881
\bibitem{druk}Drukarev E G and Levin E M 1990 \NP {\bf A511} 679; {\bf A516}
715(E)
\bibitem{sigcom}Gasser J, Leutwyler H and Sainio M E 1991 \PL {\bf B253} 252,
260
\bibitem{kusaka2}Kusaka K and Weise W 1993 {\it University of Regensburg
preprint} TPR-93-40
\bibitem{birse}Birse M C 1994 \PR {\bf C49} (in press)
\end{references}
\end{document}